\documentclass[5p]{elsarticle}

\usepackage{amsmath}
\usepackage{feynmf}
\usepackage{latexsym}
\usepackage{amssymb}
\usepackage{graphicx}
\usepackage{rotating}
\usepackage{latexsym}
\usepackage{verbatim}

\begin{document}

 \def\be{\begin{equation}}
 \def\ee{\end{equation}}
 \def\l{\lambda}
 \def\a{\alpha}
 \def\b{\beta}
 \def\g{\gamma}
 \def\d{\delta}
 \def\e{\epsilon}
 \def\m{\mu}
 \def\n{\nu}
 \def\t{\tau}
 \def\p{\partial}
 \def\s{\sigma}
 \def\r{\rho}
 \def\sl{\ds}
 \def\ds#1{#1\kern-1ex\hbox{/}}
 \def\sla{\raise.15ex\hbox{$/$}\kern-.57em}
 \def\nn{\nonumber}
 \def\bea{\begin{eqnarray}}
 \def\eea{\end{eqnarray}}
 \newcommand{\bth}{{\bf 3}}
 \newcommand{\btw}{{\bf 2}}
 \newcommand{\bon}{{\bf 1}}
 \def\QQ{{Q_Q}}
 \def\QU{{Q_{U^c}}}
 \def\QD{{Q_{D^c}}}
 \def\QL{{Q_L}}
 \def\QE{{Q_{E^c}}}
 \def\QHu{{Q_{H_u}}}
 \def\QHd{{Q_{H_d}}}
 \def\cA{\mathcal{A}}
 \def\Tr{\textnormal{Tr}}
 \def\th{\theta}
 \def\pd{\partial}
 \def\thth{\theta^2 \bar\theta^2}
 \def\sb{\bar{\s}}
 \def\psib{{\bar{\psi}}}
 \def\f{\phi}
 \def\Eps{\epsilon^{\mu\nu\rho\sigma}}
 \def\({\left(}
 \def\){\right)}
 \def\[{\left[}
 \def\]{\right]}
 \def\lb{{\bar\l}}
 \def\half{\frac{1}{2}}
 \def\la{\langle}
 \def\ra{\rangle}
 \def\numeq{n^{eq}}
 \def\D{\Delta}
 \def\cA{\mathcal{A}}
 \def\cw{\cos\th_W}
 \def\sw{\sin\th_W}
 \def\cscw{\text{cosec}\th_W}
 \def\cotw{\cot\th_W}
 \def\numeq{n^\text{eq}}
 \def\slashed{\ds}
 \def\sl{\ds}
 \def\ds#1{#1\kern-1ex\hbox{/}}
 \def\sla{\raise.15ex\hbox{$/$}\kern-.57em}
 \def\Stuckelberg{St\"uckelberg }



\begin{frontmatter}

\title{Dark Matter in Anomalous
$U(1)'$ Models with neutral mixing}

\author[label1]{Francesco
Fucito}
\author[label1]{Andrea
Lionetto}
\author[label1]{Andrea
Mammarella}

\address[label1]{Dipartimento di Fisica dell'Universit\`a di Roma ,``Tor Vergata" and
I.N.F.N.~ -~ Sezione di Roma ~ ``Tor Vergata'',\\
 Via della Ricerca  Scientifica, 1 - 00133 ~ Roma,~ ITALY}

\begin{abstract}
We study the lightest masses in the fermionic sector of an anomalous $U(1)'$ extension of the 
minimal supersymmetric standard model inspired by brane constructions. The LSP of this model 
is an XWIMP (extremely weak interaction particle) which is shown to have a relic density satisfying  
WMAP data. This computation is carried out numerically after having  adapted the DarkSUSY package to our case. 

\end{abstract}

\end{frontmatter}

\section{Introduction}

There has been much work recently to conceive an intersecting brane model with the gauge and matter 
content of the Standard Model (SM) of particle physics \cite{Marchesano:2007de,Blumenhagen:2005mu,Lust:2004ks,Kiritsis:2003mc}.
One of the features of such models is the presence of extra anomalous gauge $U(1)'$s
whose anomaly is cancelled via the Green-Schwarz mechanism. This
characterizes this class of models with respect to those which still have extra $U(1)$ gauge symmetries but
cancel the anomaly as in the SM or in the minimal supersymmetric SM (MSSM) (see \cite{Langacker:2008yv} for a review). 
The latter are inspired by (super) grand unified models or by plain extensions of the (MS)SM and their
quantum numbers with respect to the new gauge simmetries are fixed by the condition of setting to zero all the triangle
diagrams in which at least one vertex of such diagrams has a $U(1)'$ current coming from the extra gauge simmetries.
On the contrary, in brane inspired models, these quantum numbers are not fixed. We have introduced and discussed the
signatures of this model in \cite{Anastasopoulos:2008jt} and \cite{Fucito:2010dj}. Another important
issue which deserves careful scrutiny is the compatibility of this model with the WMAP data: 
this was done in \cite{FLMR} (see also \cite{Coriano:2010ww,Coriano:2010ws} for related work). The mass matrix of the fermions
uncharged with respect to the $U(1)$ of the gauge group of the SM, gets now two new contributions in this model:
one coming from the superpartner of the St\"uckelberg boson (St\"uckelino) and the other from the superpartner of the gauge boson
mediating the extra $U(1)'$. By taking some simplifying and reasonable assumptions (to be detailed later on) on the fermion masses entering the
soft supersymmetry breaking lagrangian, the LSP turns out to be the St\"uckelino. It is also easy to realize that, given the 
simplifying assumptions mentioned above, the LSP is interacting with the MSSM particles with a coupling suppressed by the inverse
of the mass of the extra gauge boson of the theory which must be at least of the order of the TeV for phenomenological reasons. 
The LSP is then an XWIMP (Extra Weakly Interacting Massive Particle), a class
of particles which have already been studied in literature \cite{Kors:2005uz,Feldman:2006wd}.
The cross section of these LSPs is too weak to give the right relic abundance. This is why one has to resort to coannihilations
with NLSPs. In our case, the cross section for the annihilations of the LSP with the NLSP and that for the coannihilations of the two species 
differ for some orders of magnitude. Once again this situation is not new in literature \cite{Klein:1999im,reliccoann,Edsjo:2003us}
but needs to be treated carefully: the two species will not decouple as far as there will be some MSSM particles to keep them in
equilibrium. Moreover these particles must be relativistic so that their abundance is enough to foster the reaction.
These points were assumed in our previous paper on this subject \cite{FLMR} to allow for a simplified treatment of the
relic abundance. In this paper we drop all simplifying assumptions and discuss the most general case,
in which the extra MSSM sector is not decoupled to the MSSM sector and the LSP is mainly a mixture of
St\"{u}ckelino and primeino, with small MSSM contribution.
We solve numerically the Boltzmann equation in this general case. To do this we have modified the DarkSUSY \cite{DarkSUSY} package to keep in account
the new interactions typical of our model. The results we obtain are qualitatively compatible with
the findings in \cite{FLMR}: there is an ample region of the parameter space which leads to a relic 
density compatible with the WMAP data. 
This is the plan of the paper:
in section \ref{sec2} we describe the neutral mixing in our model and
sketch the way in which it affects the interactions. In section \ref{sec3}
we describe how we change the DarkSUSY package and compute the relic density.

\section{Neutral mixing \label{sec2}}
Our model \cite{Anastasopoulos:2008jt} is an extension of the MSSM with an extra $U(1)$. 
The charges of the matter fields with respect to the symmetry groups are given in table 1.

  \begin{table}[h]
  \centering
  \begin{tabular}[h]{|c|c|c|c|c|}
   \hline & SU(3)$_c$ & SU(2)$_L$  & U(1)$_Y$ & ~U(1)$^{\prime}~$\\
   \hline $Q_i$   & $\bth$       &  $\btw$       &  $1/6$   & $Q_{Q}$ \\
   \hline $U^c_i$   & $\bar \bth$  &  $\bon$       &  $-2/3$  & $Q_{U^c}$
\\
   \hline $D^c_i$   & $\bar \bth$  &  $\bon$       &  $1/3$   & $Q_{D^c}$
\\
   \hline $L_i$   & $\bon$       &  $\btw$       &  $-1/2$  & $Q_{L}$ \\
   \hline $E^c_i$   & $\bon$       &  $\bon$       &  $1$     &
$Q_{E^c}$\\
   \hline $H_u$ & $\bon$       &  $\btw$       &  $1/2$   & $Q_{H_u}$\\
   \hline $H_d$ & $\bon$       &  $\btw$       &  $-1/2$  & $Q_{H_d}$ \\
   \hline
  \end{tabular}
  \caption{Charge assignment.}\label{QTable}
  \end{table} The anomalies
induced by this extension are cancelled by the GS mechanism. 
Each anomalous triangle diagram is parametrized by a coefficient $b_2^{(a)}$ (entering the lagrangian)
with the assigment:

\bea
   \cA^{(0)}:     &&\ U(1)'-U(1)'-U(1)' \rightarrow b_2^{(0)} \\
   \cA^{(1)}: &&\ U(1)'-U(1)_Y - U(1)_Y \rightarrow b_2^{(1)}\\
   \cA^{(2)}:     &&\ U(1)'-SU(2)-SU(2) \rightarrow b_2^{(2)}\\
   \cA^{(3)}:     &&\ U(1)'-SU(3)-SU(3) \rightarrow b_2^{(3)}\\
   \cA^{(4)}:    &&\ U(1)'-U(1)'-U(1)_Y \rightarrow b_2^{(4)}
\eea The mass of the extra boson is parametrized by 
$M_{V^{(0)}}=4 b_3 g_0$, where $g_0$ is the coupling of 
the extra $U(1)'$. The terms of the Lagrangian that will
contribute to our calculation are \cite{FLMR,Anastasopoulos:2008jt}:

\bea
 &&\L_{\text{St\"uckelino}}={\frac{i}{4}} \psi_S \s^\m \pd_\m \psib_S -\sqrt2 b_3 \psi_S
\l^{(0)} \nn \\ 
   &&-{\frac{i}{2\sqrt2}} \sum_{a=0}^2b^{(a)}_2  \Tr \( \l^{(a)} \s^\m
\sb^\n F_{\m
\n}^{(a)} \) \psi_S \\
   &&-{\frac{i}{2\sqrt2}} b^{(4)}_2 \[ {\frac{1}{2}} \l^{(1)} \s^\m \sb^\n
F_{\m
\n}^{(0)} \psi_S
   + (0 \leftrightarrow 1) \] +h.c.
\label{axinolagr} \nn \eea

As already said, we want to deal with the case of general neutral mixing.
The neutral mixing matrix is:
     
\be
      \( \begin{array}{c} B^{\m}\\
                          W^{3 \m} \\ C^{\m} \end{array} \) = M
            \( \begin{array}{c} A^{\m}\\
                                Z_0^{\m}\\ Z'^{\m} \end{array} \)
\label{neumix}  \ee Defining $\emph{an} \equiv g_0 ~ Q_{H_u} \frac{2 v^2}{2 M^2_{Z'}}$ we have at tree level:
\be  M= \( \begin{array}{ccc} c_W & -s_W & s_W \sqrt{g_1^2 + g_2^2}~an \\
s_W & c_W  & c_W \sqrt{g_1^2 + g_2^2}~an\\
  0 & c_W g_2+ s_W ~g_1)~an & 1    \end{array} \)
\label{neumass} \ee where $c_W$ is $cos(\theta_W)$, $s_W$ is $sin(\theta_W)$. $g_1,~g_2$ are the couplings
of the SM electro-weak $SU(2) \times U(1)$ group. The structure of this matrix leaves the electromagnetic
sector and the related quantum numbers unchanged with respect to the MSSM ones.\\
Last, we remember \cite{FLMR,Anastasopoulos:2008jt} the general form of the neutralinos mass matrix at 
tree level:
\be
    {\bf M}_{\tilde N}
     =   \(\begin{array}{cccccc}
           M_S & \sqrt{2} \frac{M_{Z'}}{2} & 0 &0 & 0 & 0 \\
           \dots & M_0 & 0  & 0 & - g_0 v_d Q_{H_u}  & g_0 v_u Q_{H_u}   \\
           \dots & \dots & M_1 & 0 & -\frac{g_1 v_d}{2} & \frac{g_1 v_u}{2} \\
           \dots & \dots& \dots & M_2 & \frac{g_2 v_d}{2} & -\frac{g_2 v_u}{2} \\
           \dots & \dots & \dots & \dots & 0 & -\m  \\
          \dots & \dots & \dots & \dots & \dots & 0
         \end{array}\) \nn ~~~~
\ee
where $M_S$ and $M_0$ are the soft masses of the st\"{u}ckelino
and of the primeino, respectively and $v_u,~v_d$ are the vevs of the Higgs fields.

\section{Numerical computations and results \label{sec3}}
Following our previous work \cite{FLMR}, in which we have studied separately the case
in which the NLSP is a bino-higgsino from that in which it is a 
wino-higgsino, we performed this general study in the same way.
The plot we will show are generated with a modified version of the
DarkSUSY package, in which we added the new fields and interactions
introduced by the anomalous extension.\\
The free parameters that we use in our numerical simulations are
the seven ones used in the MSSM-7 model: the $\m$ mass, the
wino soft mass $M_2$, the parity-odd Higgs mass $M_{A_0}$,
$tg \b$, the sfermion mass scale $m_{sf}$, the two Yukawas
$a_t$ and $a_b$. We add to this set five parameters which define 
the $U(1)'$ extension: the st\"{u}ckelino soft mass $M_S$, 
the primeino soft mass $M_0$, the $U(1)'$ charges $Q_{H_u}$, $Q_Q$, 
$Q_L$.\\
As our actual aim is to study the model without any simplification,
we can't have control on the mass gap between the LSP and NLSP,
because fixing it would require a constraint on the masses. So we
choose to let all parameters unconstrained and therefore to
collect data in the mass gap ranges 
$0\div5\%$, $5\div10\%$.
In each case we have started our study scanning the parameter space
in search of the region permitted by the experimental and theoretical constraints,
i.e. the region in which we could satisfy the WMAP data with a certain choice
of the model parameters.
After that we have numerically explored this regions to find sets of parameters that
satisfy the WMAP data for the relic density. We have found many suitable combinations
for both types of NLSP. So we have chosen some of these succesful 
models and we have computed the relic density keeping constant all but two 
parameters and plotting the results.\\
We have found regions of the parameter space in which the WMAP data
are satisfied for mass gap over $20\%$, but in the following we will only show results for the regions
$0\div5\%$ and $5\div10\%$, because they are more significant. For simplicity
we will refer to these regions as \textquotedblleft
$5\%$ region \textquotedblright and \textquotedblleft $10\%$ region\textquotedblright
respectively.\\

\subsection{General results}
In this section we want to list some results that are valid for both types
of NLSP.\\ First of all, given the constraints on the 
neutral mixing described in \cite{Langacker:2008yv}, we have obtained that
$-1 \lesssim Q_{H_u} \lesssim 1$. This implies that also in our general case the 
mixing between our anomalous LSP and the NLSP is small.\\
We have checked that there are suitable parameter space regions in which the WMAP data are satisfied
and we have found that this is true for all possible composition of our anomalous LSP.\\
We have also checked that in each region we have studied there are no divergences or unstable behaviours in our 
numerical results.\\
We have verified that the relic density is strongly dependent
on the LSP and NLSP masses and composition while it is much less dependent on the
other variables. Anyway it can be shown that there are cases 
in which the parameters not related to the LSP or NLSP can play
an important role. We will show an example of this  case in a forthcoming subsection.

\subsection{Bino-higgsino NLSP}
If the NLSP is mostly a bino-higgsino we have a two particle coannihilation.\\
We chose two sample models which satisfy the WMAP data \cite{WMAP} with mass gaps $5 \%$ and $10 \%$.
We study these models to show the dependence of the relic density from the
LSP composition and from the mass gap. To obtain these result, we have performed
a numerical simulation in which we vary only the st\"{u}ckelino and the primeino soft masses.
The results are showed in figure 1, with the conventions:
\begin{itemize}
 \item Inside the continous lines we have the region in which \newline $(\Omega h^2)_{WMAP} \sim \Omega h^2$
 \item Inside the thick lines we have the region in which \newline $(\Omega h^2-3 \s)_{WMAP}<\Omega h^2<(\Omega h^2+3 \s)_{WMAP}$
 \item Inside the dashed lines we have the region in which \newline $(\Omega h^2-5 \s)_{WMAP}<\Omega h^2<(\Omega h^2+5 \s)_{WMAP}$
 \item Inside the dotted lines we have the region in which \newline $(\Omega h^2-10 \s)_{WMAP}<\Omega h^2<(\Omega h^2+10 \s)_{WMAP}$
\end{itemize}

\begin{figure}[h!]
\includegraphics[scale=0.65]{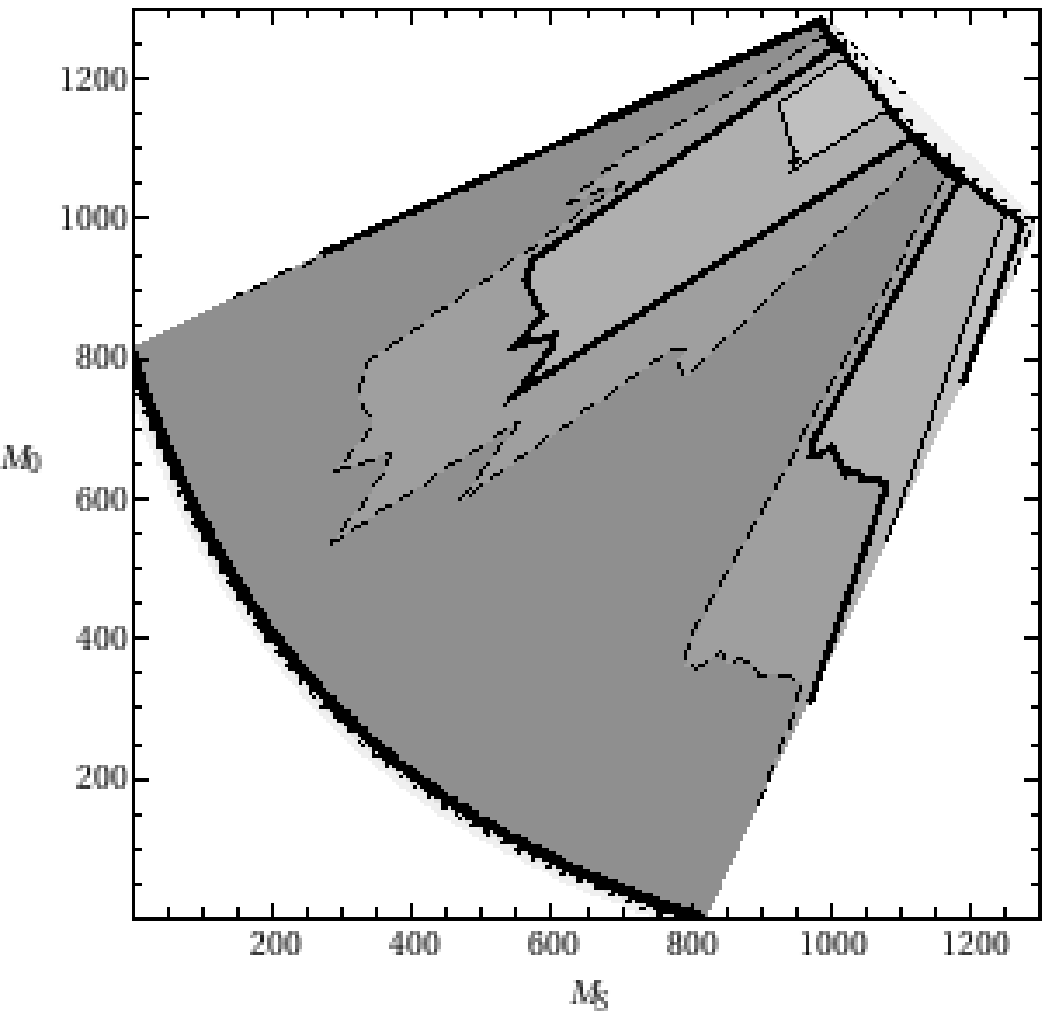}
\includegraphics[scale=0.5]{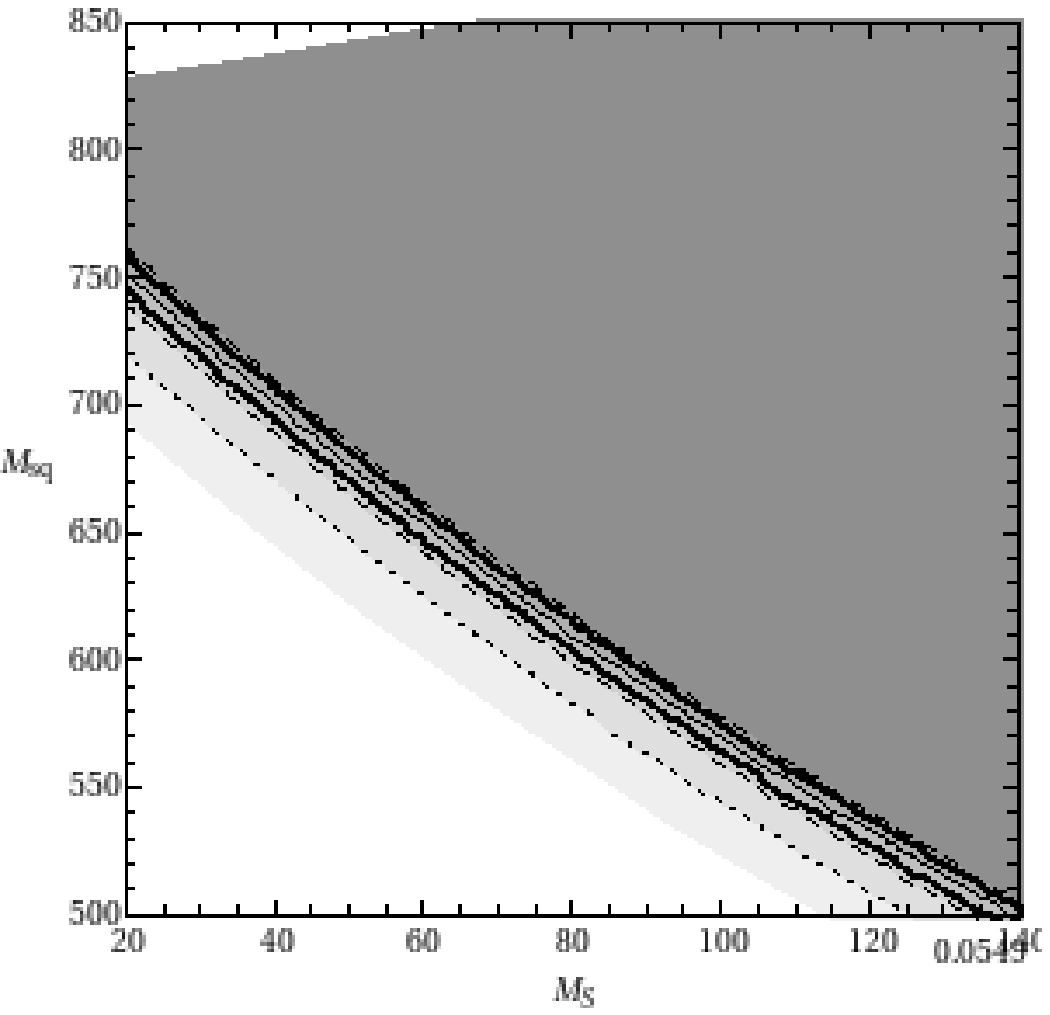}
\includegraphics[scale=0.8]{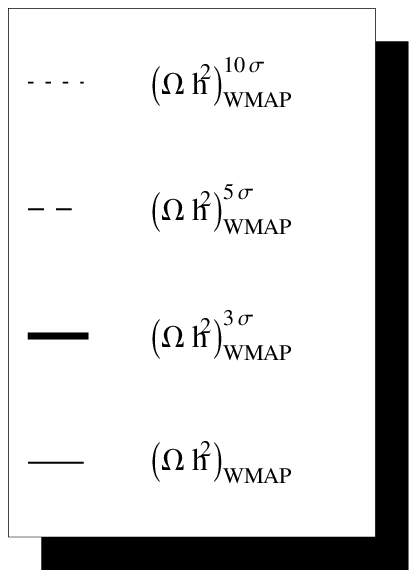}
\includegraphics[scale=0.65]{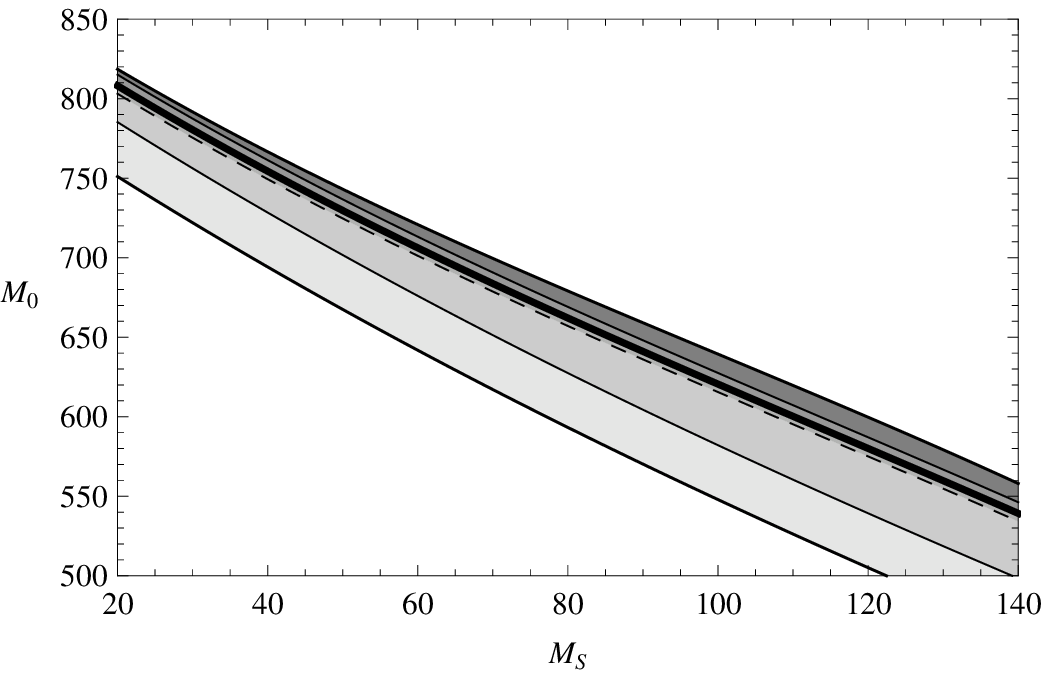}
\caption{Plot of the relic density of the LSP vs the st\"{u}ckelino
and primeino masses. The first plot shows the case of mass gap 5 $\%$. The second plot
is a zoom of the first in the region between 20-140 GeV for st\"{u}ckelino mass and 500-850 GeV
for primeino mass. The third plot shows the case of mass gap 10 $\%$ in the same region of the second plot.
All the masses are expressed in GeV}
\end{figure} Going from the region with a $5\%$ mass gap to that with a $10\%$
mass gap there is a large portion of the parameter space in which the 
WMAP data cannot be satisfied, while the regions showed in the second and third plot in
fig. 1 are similar and thus are mass-gap independent.

\subsection{Wino-higgsino NLSP}
If the NLSP is mostly a wino-higgsino we have a three particle coannihilation,
because the lightest wino is almost degenerate in mass with the lightest chargino,
so they both contribute to the coannihilations.\\
In this case we perform the same numerical calculation illustrated in the 
previous subsection. We have extensively studied a sample model with mass gap $10\%$, 
showing an example of funnel region, a resonance that occurs when $2~M_{LSP} \sim M_{A_0}$.
In our sample model $M_{LSP} \sim 330~GeV$, while $M_{A_0}\sim 630~GeV$ and this
leads to the relic density plot showed in figure 2, with the same conventions used in figure 1.
\begin{figure}[h!]
\includegraphics[scale=0.65]{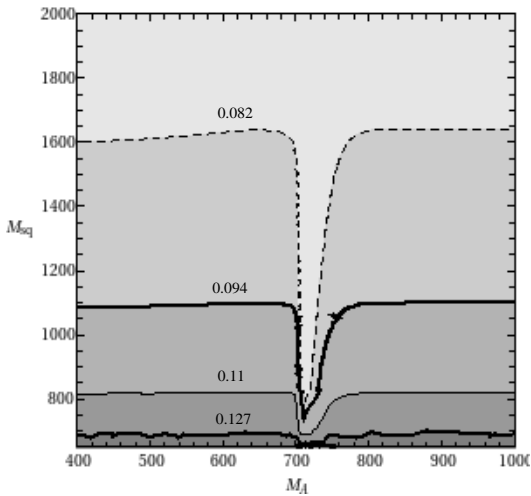} 
\caption{Example of a funnel region for a wino-like NLSP, whose mass gap with
the LSP is $10\%$}
\end{figure} So we can state that also in the case of anomalous LSP we can have a
behaviour similar to that of the MSSM, given that the LSP coannihilates with a MSSM NLSP.

\section{Conclusion \label{sec4}}
We have modified the DarkSUSY package in the routines which calculate the cross section of a given
supersymmetric particle (contained in the folder $\sim$/src/an) adding all the new interactions introduced
by our anomalous extension of the MSSM. We have also written new subroutines to calculate amplitudes that 
differ from those already contained in DarkSUSY. We have also modified the routines that generate the
supersymmetric model from the inputs, adding the parameters necessary to generate the MiAUSSM \cite{Anastasopoulos:2008jt}
and changing the routines that define the model (contained in $\sim$/src/su) accordingly. Finally
we have written a main program that lets the user choose if he wants to perform the relic density calculation
in the MSSM or in the MiAUSSM. The code of our version of the package is available contacting 
andrea.mammarella\newline @roma2.infn.it.\\
These modifications have permitted to extensively numerically explore the parameters space for an anomalous 
extension of the MSSM without restriction on the neutral mixing or on the free
parameters of the model. We have verified that our model does not lead to any divergence
or instability. \\We have also found 
sizable regions in which we can satisfy the WMAP data for mass gaps which go from $5 \%$ to beyond $20 \%$.
We have studied some specific sets of parameters for the $5\%$ and $10\%$ mass gap regions, 
showing that relatively small changes in the mass gap
can produce very important changes in the area of the regions which satisfy the experimental constraints.\\
We have also showed the presence of a funnel
region, analogous to that in the MSSM, in some region of the parameter space.\\
So we can say that a model with an anomalous LSP can satisfy all the current experimental constraints,
can show a phenomenology similar to that expected from a MSSM LSP
and can be viable to explain the DM abundance without any arbitrary constraint on its parameters.

\end{document}